# Deep V and K band photometry of the host galaxy of Haro 11

**Genoveva Micheva**[1]**, Erik Zackrisson**[2]**, Göran Östlin**[2] **and Nils Bergvall**[3]

[1] Stockholm Observatory, Department of Astronomy, Stockholm University, AlbaNova University Center, 106 91, Stockholm, Sweden

[2] Oskar Klein Centre for Cosmoparticle Physics, Department of Astronomy, Stockholm University, AlbaNova University Center, 106 91, Stockholm, Sweden

[3] Division of Astronomy & Space Physics, Uppsala university, 751 20 Uppsala, Sweden

E-mail: `genoveva@astro.su.se,ez@astro.su.se,ostlin@astro.su.se,nisse@fysast.uu.se`

**Abstract.** Previous studies of the host galaxy of *Haro 11* have suggested an extreme red color of $V-K = 4.2 \pm 0.8$ which cannot be reconciled with any normal stellar population of reasonable metallicity. We present the deepest V and K band data to date of the blue compact galaxy (BCG) *Haro 11* and derive a new $V - K$ color for the host. Our new data suggest a much more modest value of $V - K = 2.3 \pm 0.2$, which is in the same range as $V - K$ colors measured for several other BCGs. The new Haro 11 color is not abnormally red and can be attributed to an old metal-poor stellar population with a Salpeter initial mass function.

AMS classification scheme numbers: 98.52.Wz, 98.52.Ep, 98.62.Gq, 98.62.Qz

## 1. Introduction

Blue compact galaxies (BCGs) are gas-rich low-luminosity galaxies of particular interest since many of them have very high star formation rates (SFR). They are thus reminiscent of high-redshift starbursting young galaxies. This, together with their close proximity to us, makes BCGs useful test objects, suitable for gaining insight into galaxy and star formation (SF). Aside from the bright central starburst, deep photometric optical and near-infrared (NIR) studies (e.g. [1, 2, 3, 4, 6, 8]) have revealed the presence of another component of low surface brightness (LSB). The currently accepted scenario is that this LSB component is hosting the starburst and it is therefore referred to as a LSB host galaxy. The host is only visible close to the outskirts of BCGs since at small radial distances the intensity from the central starburst outshines the contribution from the LSB component. Many photometric studies have concentrated on reaching fainter magnitudes in an attempt to isolate the LSB component and characterize it. Some have suggested that the LSB component has unusual properties such as an extreme red excess in the optical/near-infrared colors which cannot be reconciled with a normal stellar population [6, 9]. Bergvall & Östlin [6] report progressively redder colors towards the outskirts of the host for a number of BCGs. Among their sample one of the biggest, most massive and most luminous BCGs is *Haro 11*, with a very high SFR of 18-20 $M_\odot \text{yr}^{-1}$ and the reddest color of $V - K = 4.2 \pm 0.8$. Current stellar evolutionary models cannot reproduce such red colors for any known normal stellar population.

Here we present the deepest observations to date for *Haro 11* in the V and K bands and derive the $V - K$ profile to test the magnitude of the red excess. In the analysis carried out by Bergvall & Östlin [6] and Zackrisson et al. [10], the BCG halo colors of the Bergvall & Östlin [6] sample can be fitted with a metal-rich stellar population following a Salpeter initial mass function (IMF). The results from this work reconcile the *Haro 11* color data point with the fit proposed by Zackrisson et al. [10].

| Haro 11 / ESO350–IG038 | | | | |
|---|---|---|---|---|
| RA (J2000) | Dec (J2000) | R | z | D |
| 00h36m52.5s | −33d33m19s | 82 Mpc | 0.020598 | ∼1′ |
| | Old data | | New data | |
| | V | K′ | V | Ks |
| Year | 1984 | 1993 | 2008 | 2005 |
| Exposure Time | 15 min | 66 min | 40 min | 60 min |
| Instrument | ESO 2.2m | ESO 2.2m | NOT 2.56m | ESO NTT 3.58m |
| ″/pix | 0.362 | 0.475 | 0.217 | 0.288 |
| FoV(″) | 178×178 | 136×136 | 462×462 | 295.2×295.2 |

**Table 1.** Observational data summary for *Haro 11*. R - distance to the target in Mpc, z - redshift, D - apparent angular diameter. The effective exposure time for new V (new K) data is 3.3 (2.4) times longer than for the old V (old K) data.

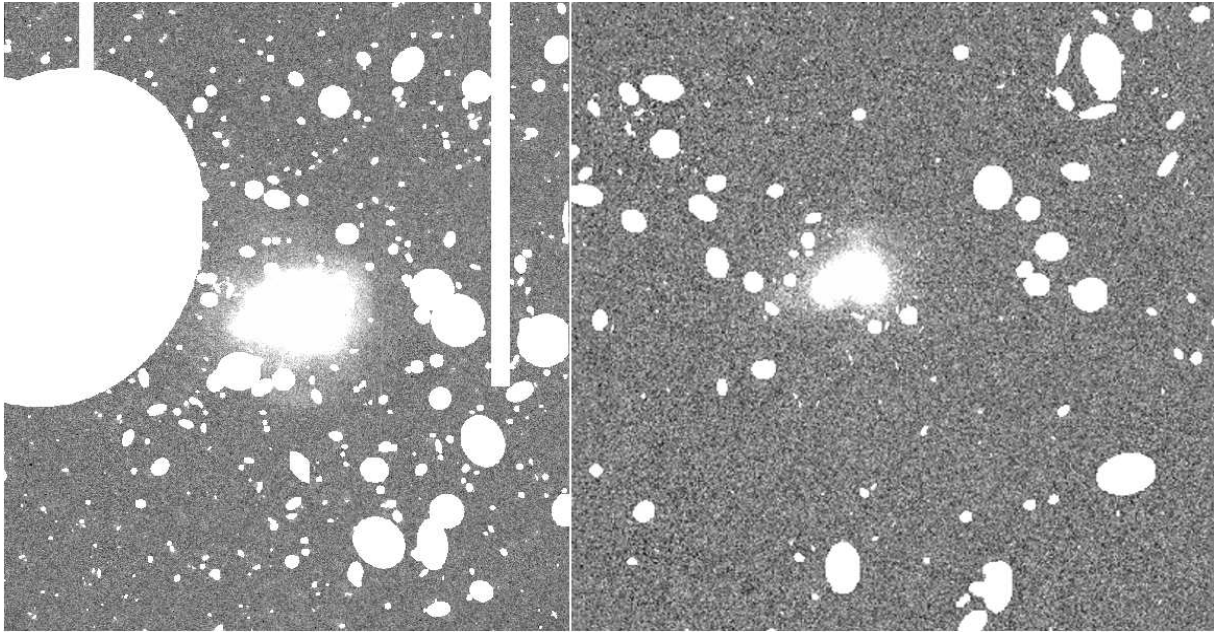

**Figure 1.** New V and K band data for *Haro 11*. Left panel: NOT MOSCA V band, 40 minute exposure, 1 chip. The image size is 2.7′× 2.3′ with a scale of 0.217 arcsec per pixel. Right panel: ESO NTT SOFI K band, 1 hour exposure. The image size is 2.3′× 2.2′ with a scale of 0.288 arcsec per pixel. Bad pixels, sources and cosmic rays in both images are masked out. The large masked out area in the V band image is a bright star. North is up, east is to the left. The larger size of the images significantly increases the area available for the estimation of the sky background.

## 2. Data reduction

Table 1 summarizes the available data used in this paper and draws a parallel to the data published by Bergvall & Östlin [6]. The V band data is obtained in 2008 at the Nordic Optical Telescope (NOT) with the MOsaic CAmera (MOSCA), which has a scale of 0.217 arcsec/pixel. The K band data is obtained at the ESO NTT 3.58m telescope with SOFI, which has a scale of 0.288 arcsec/pixel. Not taking the technical evolution of the detectors and the airmass dependency into account, 40 minutes at the NOT correspond to ∼50 minutes at the ESO 2.2 m, and 60 minutes at the ESO NTT correspond to ∼159 minutes at the ESO 2.2m.

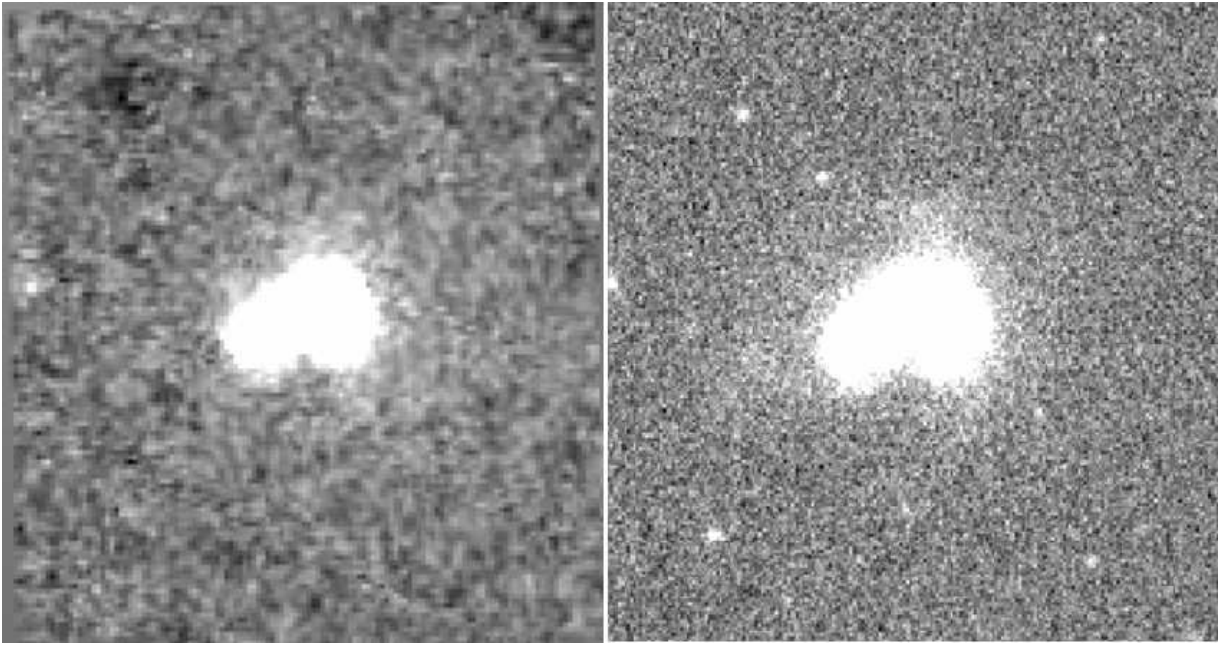

**Figure 2.** Comparison of the K band data for *Haro 11*. Left panel: the data used in the analysis of Bergvall & Östlin [6]. The image size is ∼1′×1′ with a scale of 0.475 arcsec per pixel. Right panel: the central ∼1′×1′ of the K band data used in this work. North is up, east is to the left. The new image is of better quality, with higher resolution and a flatter background.

The raw images are cleaned from bad pixels, bias subtracted in the V band, pair subtracted in the K band, trimmed, flatfielded, sky subtracted with a flat surface interpolated with a first order polynomial, calibrated with secondary standard stars in the V band and with 2MASS stars in the K band and, finally, aligned and median combined to produce the science V and K band images used for this analysis. A final sky subtraction is performed in both bands in order to remove any residual sky from the images. When performing sky subtraction we neglect the effects of dust extinction of the extragalactic background light in the outskirts of the host [12] and assume that the sky over our target can be interpolated from sky values around the target.

The surface brightness profiles are obtained by integrating in elliptical rings starting from the center of the galaxy. The parameters of the ellipse are chosen by fitting isophotes to the K band image with IRAF ELLIPSE. For robustness, different centering positions and inclinations have been tested, however there is no discernible difference in the radial color profiles obtained from integration with elliptic parameters in the range $P.A. \in (-60°, -77°)$, $e \in (0.20, 0.34)$. Once integration starts, the parameters of the elliptic rings and the step size along the major axis are kept constant. At faint isophote levels it is the uncertainty in the sky background that will largely dominate the error. We estimate this uncertainty by masking out all sources, measuring the mean intensity inside square apertures, and then measuring the standard deviation of these means. The size of the apertures is of the same order as the area of the smallest elliptic ring inside of the LSB host, at $\mu_K \sim 21.5$ mag/arcsec$^2$. This $\sigma_{sky}$ represents the uncertainty in the zero level of the background. Another error source included in the analysis is the uncertainty in the mean flux level of each ring, which is a composite error of the Poisson noise and the intrinsic flux scatter across each ring. The combination of these two error sources thus gives a rather conservative representation of the uncertainties in our results. The profiles are sampled with a step size of 5 pixels.

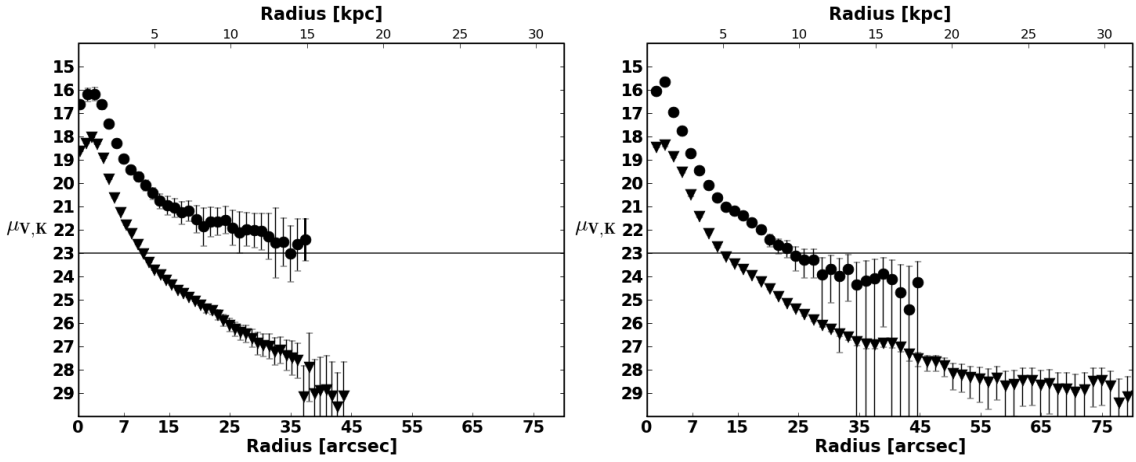

**Figure 3.** Surface brightness profiles of *Haro 11* in V (triangles) and K (circles). **Left panel:** profiles by Bergvall & Östlin [6]. The error bars represent the uncertainty in the sky level estimated to be the spread in mean values of a sample of "empty" sky regions uniformly distributed over the frames.; **Right panel:** profiles derived in this work. The error bars represent the uncertainty in the sky level together with the uncertainty of the exact mean flux value of each radius. The bumpy feature inside $r < 5$ arcsec are due to *Haro 11*'s 3 brightest knots [11], neither of which is at the center of profile integration. The horizontal line marks the faintest isophote beyond which the K band profile of an exponential disk systematically deviates from a straight line by more than 0.3 magnitudes. For comparison, the same line is marked in the left panel.

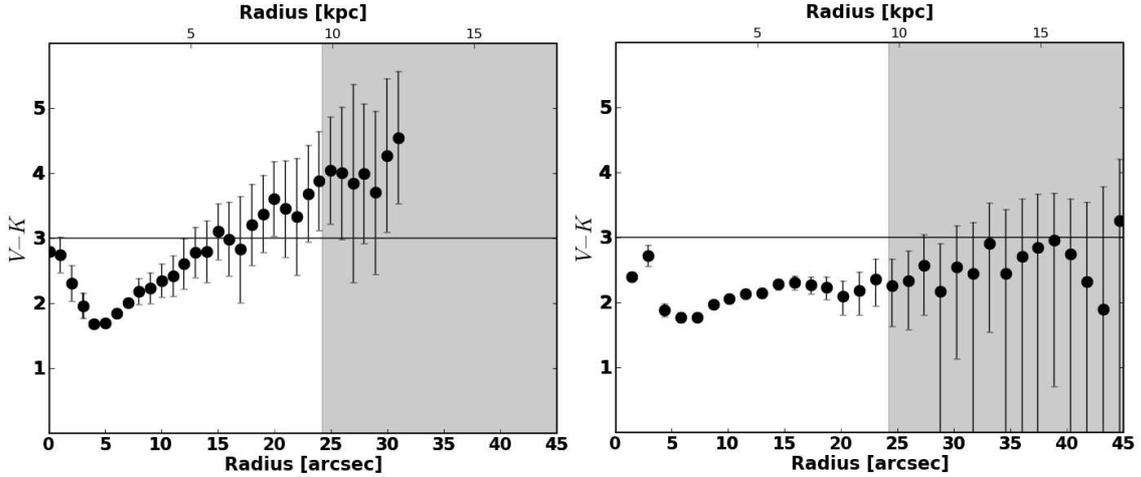

**Figure 4.** $V - K$ color profiles of *Haro 11*. **Left panel:** profile by Bergvall & Östlin [6] with a step size of 0.475 arcsec. The error bars represent the uncertainty in the zero level of the sky. No correction for dust reddening is applied. The profile becomes progressively redder with increased radius.; **Right panel:** profile derived in this work with a step size 0.548 arcsec. The error bars represent the uncertainty in the sky level together with the uncertainty of the exact mean flux value of each radius. No correction for dust reddening is applied. The bumpy features inside $r < 5$ arcsec are due to *Haro 11*'s 3 brightest knots, neither of which is at the center of profile integration. At a radius of $r \sim 14$ arcsec the color stabilizes at $V - K \sim 2.3$. The gray-shaded area corresponds to surface brightness magnitudes fainter than $\mu_K = 23$ mag/arcsec$^2$ and is not included in the calculation of the total $V - K$ color. For comparison, the same range is marked in the left panel.

## 3. Results

The intention with these observations is to test whether the red excess in the outskirts of the LSB host of *Haro 11* is indeed around $V - K \approx 4.2 \pm 0.8$ as originally suggested by Bergvall & Östlin [6]. The surface brightness profiles for V and K are presented in Fig. 3. We measure the scale length of the host population to be $\approx 2.7$ kpc in the K band by a least-squares fit of the *Haro 11* profile in the range of $\mu_K = 21.5\text{-}23.0$ mag/arcsec$^2$, which is well away from the central starburst. The extrapolated central surface brightness for this scale length is $\mu_{0,K} = 19.0$ mag/arcsec$^2$. The almost parallel behavior of the V and K profiles relative to each other hints at a flat radial color profile. Indeed, according to Fig. 4 the $V - K$ profile of the new and deeper data is clearly much flatter over the entire available radial range than the profile by Bergvall & Östlin [6]. At a radial distance of ~12 kpc from the center the color profile reaches $V - K \sim 2.3$. At the same distance, the corresponding color from Bergvall & Östlin [6] measures $V - K \sim 4$.

It should be noted that the total $V - K$ color reported in Bergvall & Östlin [6], $V - K = 4.2 \pm 0.8$, is measured out to a distance of 12.3 kpc. This is well outside of the reliability radial limit we impose on our data. Our result for the total color, $V - K = 2.3 \pm 0.2$ is measured between radii 5-10 kpc. If we instead measure the total color over the same range as Bergvall & Östlin [6], we obtain the slightly higher value of $V - K = 2.4 \pm 0.5$, which is still more than $3\sigma$ away from $V - K = 4.2$. Thus, our observations cannot verify the original detection of such an extreme red excess. The different behavior of the new and old color profiles is most likely due to the improved sky estimation in the new data. This becomes apparent especially in the K band data (Fig. 1) where the effective field of view has increased by a factor of 2.3, thus increasing the area available for sky estimation.

Assuming that a new derivation of the $B - V$ color will not significantly change the $B - V$ color, the result from this work shifts the *Haro 11* data point along the $V - K$ axis in a BVK color-color diagram. With the aid of stellar evolutionary tracks we can examine what this new V-K color implies for the outskirts of the *Haro 11* LSB host in terms of IMF and metallicity. The two scenarios presented in Zackrisson et al. [9, 10] explain the new color of the *Haro 11* host. The first is a bottom-heavy IMF ($dN/dM \propto M^{-\alpha}$ with $\alpha = 4.50$) and low to intermediate metallicity, which provides a reasonable fit to the BCGs in the Bergvall & Östlin [6] sample as well as to the new Haro 11 color measurement of $V - K = 2.3 \pm 0.2$. Zackrisson & Flynn [13] provide a detailed analysis of how this bottom-heavy IMF could contribute to the baryonic dark matter content of the Universe and impose constraints on the composition of such red halo structures. Zackrisson et al. [9, 10] offer an additional scenario which involves a Salpeter IMF with low to intermediate metallicity. For some BCGs in the sample an intermediate metallicity of $Z = 0.004\text{-}0.008$ is inconsistent with the interstellar gas metallicity measured in the central starburst, $Z \leq 20\% Z_\odot$ [6]. For the new Haro 11 color presented in this paper, however, a fit with stellar evolutionary tracks infers a halo metallicity of $Z = 0.001$ for an old stellar population with a normal IMF. Hence, the metallicity indicated by the color is consistent with that measured from emission-line ratios, and an anomalous stellar population is no longer required to explain the properties of the Haro 11 host galaxy.

**Acknowledgments**

This work includes observations made with the ESO NTT telescope at the La Silla Observatory under program ID 075.B-0220 and with the Nordic Optical Telescope, operated on the island of La Palma jointly by Denmark, Finland, Iceland, Norway, and Sweden, in the Spanish Observatorio del Roque de los Muchachos of the Instituto de Astrofisica de Canarias.
Special thanks to Tapio Pursimo for performing the V band observations, and to Robert Cumming and Angela Adamo for useful discussions.

EZ acknowledges a research grant from the Swedish Royal Academy of Sciences.
GÖ and NB acknowledge support from the Swedish Research Council.

**References**

[1] Loose, H.-H. & Thuan, T.X. 1986, ApJ, **309** 59
[2] Doublier, V., Comte, G., Petrosian, A., Surace, C., and Turatto, M. 1997, A&AS **124** 405
[3] Papaderos, P., Loose, H.-H., Thuan, T.X., and Fricke, K.J. 1996, A&AS **120** 207
[4] Cairós, L.M., Vílchez, J.M., Pérez, J.N.G., Páramo, J.I., and Caon, N. 2001, ApJS **133** 321
[5] Bergvall, N., Marquart, T., Persson, C., Zackrisson, E., Östlin, G. 2005, Multiwavelength Mapping of Galaxy Formation and Evolution **355**
[6] Bergvall, N. & Östlin, G. 2002, A&A **390** 891
[7] Marigo, P., Girardi, L., Bressan, A., Groenewegen, M. A. T., Silva, L., & Granato, G. L. 2008, A&A **482** 883
[8] Noeske, K.G., Papaderos, P., Cairós, L.M., and Fricke, K.J. 2003, A&A **410** 481
[9] Zackrisson, E., Bergvall, N., Östlin, G., Micheva, G., and Leksell, M. 2006, ApJ **650** 812
[10] Zackrisson, E., Micheva, G., Bergvall, N., and Östlin, G. 2009, these proceedings, (Preprint: arXiv0902.4695)
[11] Cumming, R., Östlin, G., Marquart, T., Fathi, K., Bergvall, N., and Adamo, A. 2009, these proceedings, (Preprint: arXiv0901.2869)
[12] Zackrisson, E., Micheva, G., and Östlin, G. 2009, submitted to MNRAS
[13] Zackrisson, E., & Flynn, C. 2008, ApJ **687** 242